\documentclass[sigconf]{acmart}

\usepackage{latexsym}
\usepackage{graphicx}
\graphicspath{{./images/}}
\usepackage{booktabs} 
\usepackage{color}  
\usepackage{amsmath}  
\usepackage{subcaption}
\usepackage{caption}
\usepackage{tikz}
\usepackage{colortbl} 
\usepackage{framed}
\usepackage{multirow}
\usepackage{multicol}
\usepackage{hyperref}
\usepackage{url}
\usepackage{balance}
\usepackage{verbatim}
\usepackage{cancel}

\newcommand{\utterance}[1]{\textit{``#1''}}
\newcommand{\phrase}[1]{\textit{`#1'}}
\newcommand{\old}[1]{}

\newcommand{\onecolscale}{0.21}

\newenvironment{Snugshade}[1][236,236,236]{
    \setlength{\itemsep}{0pt}
     \setlength{\parsep}{0pt}
     \setlength{\topsep}{0pt}
     \setlength{\partopsep}{0pt}
     \setlength{\leftmargin}{1.5em}
     \setlength{\labelwidth}{0em}
     \setlength{\labelsep}{0em} 
\setlength{\parskip}{0pt}
    \definecolor{shadecolor}{RGB}{#1}%
    \begin{snugshade}
}{%
    \end{snugshade}%
}

\setcopyright{none}

\acmConference[SIGIR'19]{ACM SIGIR Conference}{July 2019}
{Paris, France}
\acmYear{2019}
\copyrightyear{2019}

\acmArticle{4}
\acmPrice{15.00}

\newcommand\method[1]{{\sf\small{#1}}}
\newcommand{\quest}{\mbox{\method{QUEST}}}
\newcommand{\bfs}{\mbox{\method{BFS}}}
\newcommand{\shortestPath}{\mbox{\method{ShortestPaths}}}
\newcommand{\paralex}{\mbox{\method{PARALEX}}}

\newcommand{\bfsat}[1]{\mbox{\bfs@$#1$}}
\newcommand{\questat}[1]{\mbox{\quest@$#1$}}

\newcommand{\APatk}[1]{\mbox{\AP@${#1}$}}
\newcommand{\MRRatk}[1]{\mbox{\MRR@${#1}$}}
\newcommand{\Hitatk}[1]{\mbox{\Hit@${#1}$}}

\newcommand{\mycaption}[1]{\caption{\normalfont{#1}}}
\newcommand{\myparagraph}[1]{\vspace{0.3\baselineskip}\noindent{\textbf{#1}}.~}
\newcommand{\omt}[1]{}

\begin{document}

\title{Answering Complex Questions by Joining\\Multi-Document Evidence with
Quasi Knowledge Graphs}

\author{Xiaolu Lu}
\authornote{Part of this work was done during the author's internship
	at the MPI for Informatics.}
\affiliation{%
  \institution{RMIT University, Australia}
}
\email{xiaolu.lu@rmit.edu.au}

\author{Soumajit Pramanik}
\affiliation{%
	\institution{MPI for Informatics, Germany}
}
\email{pramanik@mpi-inf.mpg.de}

\author{Rishiraj Saha Roy}
\affiliation{%
	\institution{MPI for Informatics, Germany}
}
\email{rishiraj@mpi-inf.mpg.de}

\author{Abdalghani Abujabal}
\affiliation{%
	\institution{Amazon Alexa, Germany}
}
\email{abujabaa@amazon.de}

\author{Yafang Wang}
\affiliation{%
	\institution{Ant Financial Services Group, China}
}
\email{yafang.wyf@antfin.com}

\author{Gerhard Weikum}
\affiliation{%
	\institution{MPI for Informatics, Germany}
}
\email{weikum@mpi-inf.mpg.de}

\renewcommand{\shortauthors}{X. Lu et al.}

\newcommand{\squishlist}{
 \begin{list}{$\bullet$}
  { \setlength{\itemsep}{0pt}
     \setlength{\parsep}{1pt}
     \setlength{\topsep}{1pt}
     \setlength{\partopsep}{0pt}
     \setlength{\leftmargin}{1.5em}
     \setlength{\labelwidth}{1em}
     \setlength{\labelsep}{0.5em} } }

\newcommand{\squishend}{
  \end{list}  }

\newcommand{\comm}[1]{}

\begin{abstract}
Direct answering of questions that involve
multiple entities and relations is a challenge for text-based QA.
This problem is most pronounced when answers can be found only by
joining evidence from multiple documents. Curated knowledge graphs (KGs) 
may yield good answers, but are limited by their inherent
incompleteness and potential staleness. 
This paper presents
QUEST, a method that can
answer complex questions directly from textual sources on-the-fly, by
computing similarity joins over partial results from different documents.
Our method is completely unsupervised, avoiding training-data bottlenecks
and being able to cope with rapidly evolving ad hoc topics
and formulation style
in user questions.
QUEST builds a noisy quasi KG with node and edge weights, consisting of
dynamically retrieved entity names and relational phrases. 
It augments this graph with types and semantic alignments,
and computes the best answers by
an algorithm for Group Steiner Trees. 
We evaluate QUEST on benchmarks of complex
questions, and show that it substantially outperforms state-of-the-art
baselines.
\end{abstract}

%
%




\maketitle

\section{Introduction}
\label{sec:introduction}

\subsection{Motivation}
%
Question answering (QA)
over the Web and derived sources of knowledge,
has been well-researched 
\cite{kwok01scaling,fader2014open,chen2017reading}.
Studies show that
many
Web search queries 
have the form of questions~\cite{white2015questions};
and this fraction is increasing as voice search becomes
ubiquitous~\cite{guy2018characteristics}.
The focus of this paper is on providing {direct answers} to
fact-centric ad hoc
questions,
posed in
{natural
language},
or in  {telegraphic
form}~\cite{sawant2013learning,joshi2014knowledge}. 
Earlier approaches for such QA, up to when IBM Watson~\cite{watson2012}
won the
Jeopardy! quiz show, have mostly tapped into textual sources
(including Wikipedia articles)
using passage retrieval and other techniques
~\cite{ravichandran2002learning,voorhees2005trec}.
In the last few years, the paradigm of translating questions
into formal queries over structured knowledge graphs (KGs), also
known as knowledge bases (KBs),
and databases (DBs), including Linked Open Data,
has become prevalent~\cite{DBLP:conf/rweb/UngerFC14,
DBLP:journals/kais/DiefenbachLSM18}.

QA over structured data
translates the terms in a question into 
the vocabulary of the underlying KG or DB: entity names, semantic types,
and predicate names for attributes and relations.
State-of-the-art systems~\cite{DBLP:conf/emnlp/BerantCFL13,bast2015more,
DBLP:conf/acl/YihCHG15,
abujabal2017automated,fader2014open}
perform well for simple questions
that involve a few predicates around a single target entity 
(or a qualifying entity list).
However, for complex questions that refer to multiple entities and
multiple relationships between them, 
the question-to-query translation is very challenging and becomes
the make-or-break point.
As an example, consider the question:
%
\begin{Snugshade}
\noindent \utterance{footballers of African descent who played in
the FIFA 2018 final
and the Euro 2016 final?}
\end{Snugshade}
A high-quality, up-to-date KG would have answers like
{\phrase{Samuel Umtiti}, \phrase{Paul Pogba} or \phrase{Blaise Matuidi}.
However, this works only with a perfect mapping of question terms
onto KG-predicates like {\small\tt bornIn}, {\small\tt playedFor},
{\small\tt inFinal}, etc.
This strong assumption is rarely satisfied for such complex questions.
Moreover, if the KG misses some of the relevant pieces, for example,
that a football team played in a final (without winning it), 
then the entire query will fail.



\vspace*{-0.3cm}
\subsection{State of the Art and its Limitations}
\myparagraph{QA over KGs}
State-of-the-art work on QA over KGs has several
critical limitations:
(i) The question-to-query translation is brittle and tends to
be infeasible for
complex questions.
(ii) Computing good answers depends on the completeness and
freshness of the underlying KG, but no KG covers everything and keeps up
with the fast pace of real-world changes.
(iii) The generation of good queries is tied to
a specific language (usually 
English)
and style (typically full interrogative sentences), and
does not generalize to
arbitrary languages and language registers.
(iv) Likewise, the translation procedure is sensitive to the choice of the
underlying KG/DB source, and cannot handle ad hoc choices of
several sources
that are seen only at QA-execution time.

%
%

\myparagraph{QA over text}
State-of-the-art methods in this realm 
face major obstacles:
(i) To retrieve relevant passages for answer extraction,
all significant question terms must be matched in the {\em same document},
ideally within short proximity.
For example, if a QA system finds
players in the 2018 FIFA World Cup Final and the UEFA Euro Cup 2016 Final
in \textit{different} news articles, and information on
their birthplaces from 
Wikipedia, there is no single text passage for proper answering.
(ii) The alternative of fragmenting complex questions into
simpler sub-questions
requires
syntactic decomposition patterns that break with ungrammatical
constructs, and also a way of stitching sub-results together.
Other than for special cases like temporal modifiers in questions,
this is beyond the scope of today's systems.
(iii) Modern approaches that leverage deep learning critically rely
on training data, which is not easily available for complex questions,
and have concerns on interpretability.
(iv) Recent works on text-QA considered scenarios where a question
is given
with a specific set of documents that contain
the answer(s).

This paper overcomes these limitations
by providing
a novel unsupervised method
for combining answer evidence from 
multiple documents retrieved dynamically, joined together
via KG-style relationships automatically gathered from text sources.

\begin{figure} [t]
  \centering
   \includegraphics[width=0.8\columnwidth]
   {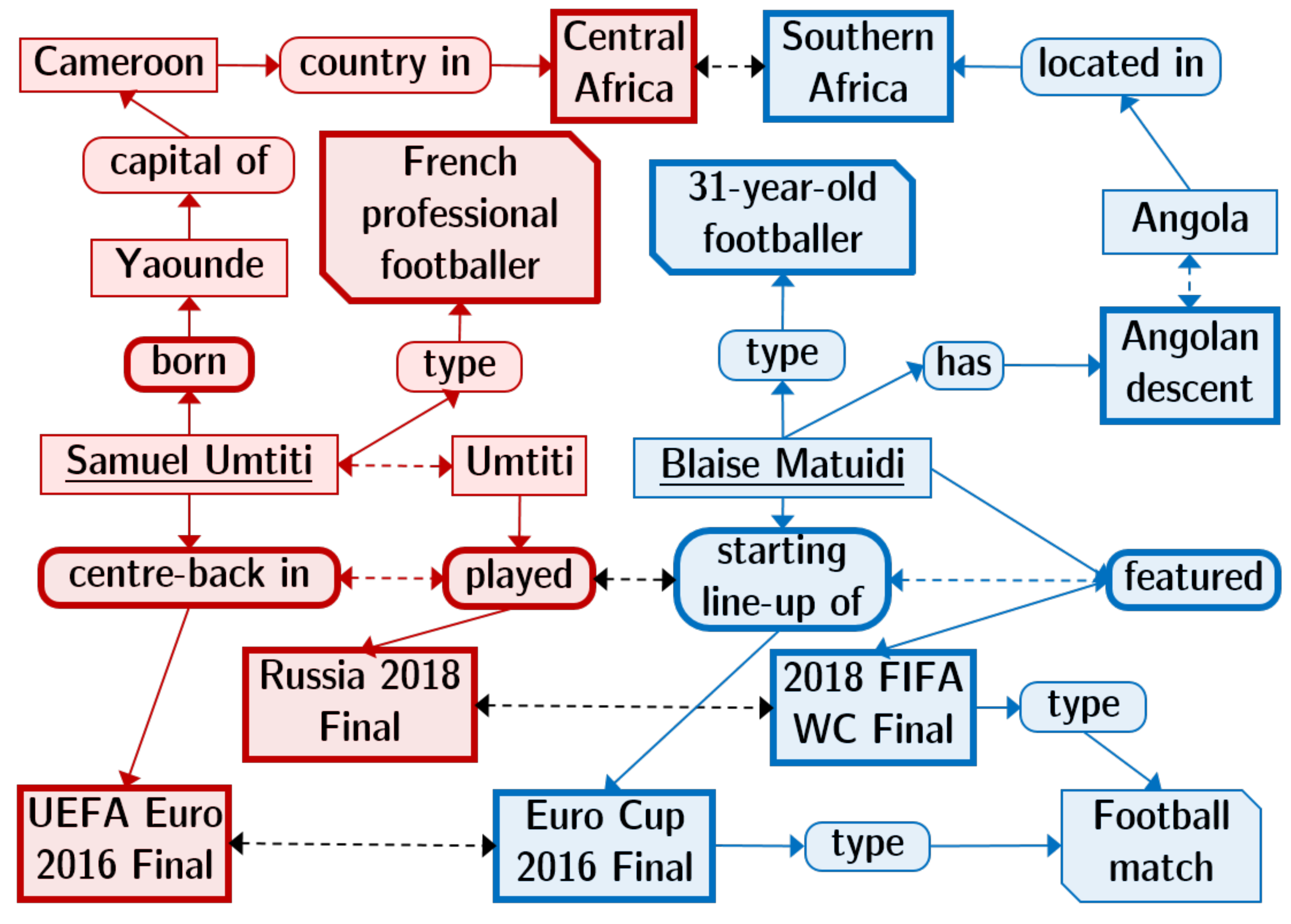}
      \mycaption{A quasi KG for our running example question.}
      \label{fig:quasiKG-example-new}
      \vspace*{-0.4cm}
\end{figure}

\subsection{Approach and Contribution} 
%
We present a method and system, called {\quest} (for ``QUEstion answering
with Steiner Trees''),
that taps into text sources for answers, but also integrates considerations
from the KG-QA (also referred to as KB-QA) paradigm.
{\quest} first constructs an ad hoc, noisy knowledge graph by dynamically
retrieving
question-relevant text
documents
and running Open Information Extraction
(Open IE)~\cite{mausam2016open}
on them to produce subject-predicate-object (SPO) triples. In contrast to
a curated KG (like YAGO or Wikidata), these triples
contain names and phrases rather than canonicalized entities
and predicates,
and hence exhibit a high degree of noise. Thus, we additionally compute
edges that
connect potentially synonymous names and phrases.
We refer to the resulting graph as a {\em quasi KG},
which captures combined cues from many documents and is then treated
as the knowledge source for the QA algorithm
(example in Fig.~\ref{fig:quasiKG-example-new} for the footballer question,
with bidirectional dashed edges denoting potential synonymy among
the nodes).

Good answers among the nodes of the 
quasi-KG
should be well-connected with all nodes that (approximately) match
the phrases from the input question. We refer to these matching nodes
as {\em cornerstones} (nodes with thick borders in
Fig.~\ref{fig:quasiKG-example-new}).
This criterion can be cast into
computing Group Steiner Trees (GST)
with the cornerstones as
terminals~\cite{DBLP:journals/jal/GargKR00}.
All non-terminal nodes of the trees are candidate answers.
This computation is carried out over a weighted graph, with weights
based on
matching scores and extraction
confidences. Finally, answers are filtered and ranked by whether they are
compatible
with the question's lexical answer type and other criteria.
In Fig.~\ref{fig:quasiKG-example-new}, all red nodes and edges, and
all blue nodes and edges, constitute two
GSTs, respectively yielding the answers \phrase{Samuel Umtiti} and
\phrase{Blaise Matuidi} (underlined).
Unlike most QA systems where correct
entity and relation linking are major bottlenecks for success,
{\quest} \textit{does not need
any explicit disambiguation} of the question concepts, and
instead harnesses
the effect that GSTs themselves establish a common context for
ambiguous names.
Thus, finding a GST serves as a \textit{joint disambiguation} step for
relevant entities,
relations, and types, as different senses of polysemous concepts
are unlikely
to share several inter-connections. Notably, the Steiner tree provides
\textit{explainable} insights into how an answer is derived.

It is the nature of ad hoc Web questions that dynamically retrieved
documents contain a substantial amount of uninformative and
misleading content.
Instead of 
attempting to
perfectly clean this input
upfront, our rationale is to
cope with this noise
in the answer computation rather than through
tedious efforts on
entity disambiguation and 
relation
canonicalization 
in each step.

GST algorithms have been used for keyword search over
relational
graphs~\cite{DBLP:series/synthesis/2010Yu,bhalotia2002keyword,
coffman2014empirical,chanial2018connectionlens}, but 
they work for
a simple class
of keyword queries with the sole condition of nodes being related.
For QA,
the problem is much more difficult as the input is a full question
that contains multiple conditions with different entities and predicates.

\myparagraph{Contribution}
%
The salient points of this work are:
\squishlist
\item {\quest} is a novel 
method
that 
computes direct answers to
{complex questions} by 
dynamically tapping arbitrary text sources and joining sub-results
from multiple documents.
In contrast to 
neural QA methods
that rely on substantial amounts of training data,
{\quest} is {unsupervised},
avoiding training bottlenecks and the potential bias towards
specific benchmarks.
\item {\quest} combines the versatility of text-based QA 
with
{\em graph-structure awareness} of KG-QA, overcoming the problems of
incompleteness, staleness, and brittleness of QA over KGs alone.
\item We devise advanced graph algorithms for computing answers from
noisy text-based entity-relationship graphs.
\item Experiments show the viability of our method and
its superiority over 
state-of-the-art baselines.
To facilitate comparison and reproducibility,
an online \textbf{demo}, and
all \textbf{data, code}, and \textbf{results}
from this work
are publicly available
at the following URL:
\def\UrlFont{\bfseries}
\url{http://qa.mpi-inf.mpg.de/quest/}.
\squishend

\section{System Overview}
\label{sec:overview}


\myparagraph{Complex questions}
Our emphasis is on complex questions that refer to multiple entities and
relationships. 
There are other notions of complex questions, for example, those requiring
grouping, comparison, and aggregation,
or when the question involves negations.
These are not considered in this paper.

\myparagraph{Answering pipeline} {\quest} processes questions
in two phases: 
(1) on-the-fly construction of the quasi KG for the question, and,
(2) the graph algorithm for computing ranked answers.
Together, these phases comprise the following
five steps:
{\setlength{\leftmargin}{2.5em}
\squishlist
\item[(1a)] Retrieving question-relevant documents from an open corpus
(e.g., the entire Web),
\item[(1b)] Extracting proximity-based SPO
triples from these documents using Open IE techniques,
\item[(1c)] Building a noisy quasi KG from these triples,
\item[(2a)] Computing
GSTs on the 
quasi KG to derive candidate answers, 
\item[(2b)] Filtering and scoring candidates to produce 
ranked answers.
\squishend
}
{\setlength{\leftmargin}{1.5em}

Step (1a), document retrieval, is performed using any Web search engine
or an IR system.
Working
with a reasonably-sized pool of pseudo-relevant documents ensures that 
the subsequent
graph algorithm is computationally tractable.
Preprocessing of documents includes
part-of-speech (POS) tagging, named entity recognition (NER),
and lightweight coreference resolution
by linking personal and possessive pronouns 
like
\textit{he, she, him, her, his}, and \textit{hers},
to the nearest
named entity in the preceding text. 
Sec.~\ref{sec:graphconstruction} and~\ref{sec:graphalgorithm}
give details on steps (1b) and (1c) for graph construction
and steps (2a) and (2b) for the graph algorithm, respectively.

\section{Graph Construction}
\label{sec:graphconstruction}


\subsection{Extracting SPO Triples}
\label{subsec:triples}



%
Answering complex questions often requires evidence from
multiple documents. 
For our running example,
the evidence needs to 
comprise \textit{born in Africa,
being a footballer,
playing in the UEFA Euro 2016 final},
and \textit{playing in the FIFA World Cup 2018 final}.
It is unlikely
that all these cues can be found in a single
document.
Therefore, we first retrieve a pool of relevant documents
from a search engine (Google or Bing) using the entire question as a
keyword query.
To identify cues in the matching documents
and to join them
across multiple
documents, we apply 
Open IE~\cite{mausam2016open,kadry2017open} to
extract SPO 
triples.

Popular tools for Open IE include 
Stanford 
OpenIE
~\cite{angeli2015leveraging},
OpenIE $5.0$
(previously 
ReVerb)
~\cite{mausam2016open}, and
ClausIE~\cite{del2013clausie}, 
but each comes with limitations.
Stanford OpenIE focuses on precision, and produces correct 
but relatively few triples, thus
losing cues from long sentences.
For example, consider the opening sentence from Wikipedia's article
on Umtiti:
\utterance{Samuel Yves Umtiti is a French professional footballer
	who plays
as a centre-back for Spanish club Barcelona and
the French National Team.}
Stanford's extractor misses the noun-mediated relation
\phrase{centre-back for},
and the information about playing for the French National Team.
While OpenIE $5.0$
and ClausIE incorporate better support for such dependent clauses,
they often produce very long objects from
complex sentences,
making it difficult to align them across different triples
and rendering subsequent graph construction infeasible
(e.g. \phrase{a French professional footballer who plays as a centre-back
	for Spanish club Barcelona and the French National Team}
and 
\phrase{as a centre-back for the French National Team} are objects
from OpenIE and ClausIE from the sentence above).
%
Therefore, we devised our own
(recall-oriented) IE method with judicious
consideration to
{\em phrase ordering} and 
{\em term proximities}.
In contrast to the other tools, our method
produces more noisy triples, but these
are taken care of in the subsequent
steps (see Sec.~\ref{sec:graphalgorithm}).

\myparagraph{Pattern-based extraction}
We perform
POS tagging and 
NER on each sentence of
the retrieved documents, and subsequently treat the
constituent words of named entities as single tokens.
Based on these annotations, we extract the following SPO triples
respecting phrase \textit{ordering}: 
\squishlist
\item {\bf Verb-phrase-mediated triples:} 
$(X,V,Y)$ for named entities (NE) or noun phrases (NP) $X$ and $Y$
such that the sentence has the form
``$\ldots X \ldots V \ldots Y \ldots$'', where
$V$ is of the POS pattern  {\small\tt verb} (e.g., ``plays'') 
or {\small\tt verb+preposition} (e.g., ``plays for''),
and no other verb appears in the text span from $X$ to $Y$.
\item  {\bf Noun-phrase-mediated triples:} 
$(X,N,Y)$ for NEs or NPs $X$ and $Y$
such that the sentence has the form
``$\ldots X \ldots N \ldots Y \ldots$'', where
$N$ is of the POS pattern {\small\tt noun+preposition}
(e.g., ``centre-back for''),
and no other phrase of this POS pattern appears in the text span from
$X$ to $Y$ (rules from Yahya et al.~\cite{yahya2014renoun}).
\squishend
Auxiliary verbs (\textit{is, can, may, ...}) are not considered.
Sample triples from {\quest}'s noisy extractor for
the previous example:
\textit{(Samuel Umtiti, centre-back for, Spanish club Barcelona),
(Samuel Umtiti, centre-back for, French National Team)} (both correct),
and
\textit{(French professional footballer, plays as,
	Spanish club Barcelona)}
(noisy).

\myparagraph{Proximity-based scoring}
To associate an SPO triple with a confidence score,
we use pairwise distances between
the triple's parts ($S$, $P$, or $O$) in the document
where it stems from.
We define the distance $d$ between two items as
the number of intruding words plus one (to avoid zero distances),
and the score is set to $1/d$.
This captures the intuition that closer two parts are in text,
higher is their score.
As a result, unlike conventional 
tools~\cite{angeli2015leveraging,mausam2016open},
S-P and P-O pairs are
allowed to take different scores. 
When two items co-occur in
more than one sentence $\{\mathcal{S}_i\}$, 
we use the
\textit{sum} of inverses of their distances in these sentences
$\{d(\mathcal{S}_i)\}$
as the score ($= \sum_{\mathcal{S}_i} 1 / d(\mathcal{S}_i)$), thereby 
\textit{leveraging redundancy of evidence}.

\subsection{Building the Quasi KG}
\label{subsec:graph}

\myparagraph{Quasi KG} 
The quasi KG consists of:
\squishlist
\item nodes corresponding to the $S$, $P$, and $O$ arguments of
the extracted triples,
\item type nodes corresponding to the $S$ and $O$ arguments,
\item directed edges connecting $S$ to $P$ and $P$ to $O$ from
the extracted triples above,
\item directed (bidirectional) edges connecting nodes with equal
or highly similar names or phrases, and,
\item directed edges connecting $S$ or $O$ nodes to their
types.
\squishend

The quasi KG is 
the key asset to aggregate evidence from multiple documents. 
Typically, an $O$ node from one triple would have an edge to an $S$
node from
another triple if they refer to the same entity, or they would be
the same node
in case they match exactly. 
Note, however, that the graph contains only surface forms; there is no
canonicalization of entity names or phrases for $S$/$O$ nodes --
hence our
terminology of \textit{quasi} KGs.
For example, 
`Pogba', `Paul Pogba' and `Paul Labile Pogba'
co-exist
as different nodes in the graph,
but could later be identified as synonyms for the same entity.
%
This notion of similarity-based equivalence also holds for
nodes that represent $P$ arguments by phrases such as \phrase{born in}
and \phrase{comes from},
and will be harnessed in subsequent steps.
%
%
%
%
Fig.~\ref{fig:quasiKG-example-new} shows an
example of a noisy quasi KG that contains
evidence for answers of our running example.
Bidirectional 
dashed edges denote potential synonymy between nodes. 

\myparagraph{Types of nodes} 
$S$, $P$, and $O$ arguments of extracted triples become
individual nodes. 
Nodes are typed based on their semantic roles:
\squishlist
\item {\bf Entity nodes} are all nodes corresponding to $S$ or $O$
arguments
(regular rectangular nodes in Fig.~\ref{fig:quasiKG-example-new}, e.g.
\phrase{Samuel Umtiti}).
\item {\bf Relation nodes} are all nodes corresponding to $P$ arguments
(rectangular nodes with rounded corners, e.g. \phrase{centre-back in}).
\item {\bf Type nodes}, explained below, are added by
inferring semantic types
for entity nodes
(rectangular nodes with snipped corners,
e.g. \phrase{French professional footballer}).
\squishend
The distinction into entity and relation nodes is important for
answer extraction later,
as relations rarely ever serve as final answers.
Usually, KGs represent relations as edge labels; our choice of
having explicit
relation nodes is better suited for the graph algorithms
that {\quest} runs
on the quasi KG.

{\quest} creates
\emph{type nodes} connected to entity
nodes 
%
via \textit{Hearst patterns}~\cite{hearst1992automatic} 
(e.g., \textit{$NP_1$ such as \{$NP_2$,\}*} with POS tag $NP$
for noun phrases)
applied to sentences of the retrieved documents
(e.g., ``\textit{footballers} such as Umtiti, Matuidi and Pogba'').
%
As an alternative or complement,
{\quest} can also
use type information from curated KGs
by
mapping the names in the
entity nodes to KG entries by
named entity disambiguation (NED) methods. 
Like entities and relations, types are not canonicalized to any taxonomy.
We found that even such free-form text extractions for
entity typing is a valuable asset for QA, as substantiated
by our graph ablation experiments.

\myparagraph{Types of edges} 
The quasi KG contains four types of edges:
\squishlist
\item {\bf Triple edges} connect two adjacent arguments of
the extracted triples, i.e., $S$ and $P$ arguments, and
$P$ and $O$ arguments from the same triple
(entity-relation solid edges,
e.g., between \phrase{Blaise Matuidi} and \phrase{featured}
and, \phrase{featured} and \phrase{2018 FIFA WC Final}).
\item {\bf Type edges} connect entity nodes and
their corresponding type nodes
(e.g., between \phrase{Blaise Matuidi} and \phrase{type},
and \phrase{type} and \phrase{31-year-old footballer}).
\item {\bf Entity alignment edges} connect potentially equivalent
entity nodes, that is, with sufficiently high similarity (dashed edges,
e.g., 
between \phrase{Samuel Umtiti} and \phrase{Umtiti}). 
\item {\bf Relation alignment edges} connect potentially 
synonymous relation nodes (dashed edges, e.g.,
between \phrase{played} and \phrase{starting line-up of}). 
\squishend
%
%
%
%
To identify node pairs for alignment edges, we can harness
resources like
\textit{entity-mention dictionaries}
~\cite{hoffart2011robust,DBLP:conf/lrec/SpitkovskyC12},
\textit{paraphrase databases}
\cite{pavlick2015ppdb,DBLP:conf/emnlp/GrycnerW16},
and
\textit{word/phrase embeddings}~\cite{mikolov13distributed,
DBLP:conf/emnlp/PenningtonSM14}.
%



\myparagraph{Node weights}
Node weights are derived from
similarity scores with regard
to tokens in the 
input question.
For entity nodes, we use 
thresholded similarities from entity-mention dictionaries
as explained in Sec.~\ref{subsec:corpora}.
For type nodes and relation nodes, the similarity of the node label
is with regard to 
the highest-scoring
question token (after stopword removal).
In
{\quest},
this similarity 
is computed using 
word2vec~\cite{mikolov13distributed},
GloVe~\cite{DBLP:conf/emnlp/PenningtonSM14},
or BERT~\cite{devlin2018bert} embeddings.

\myparagraph{Edge weights} 
%
For \textit{triple edges}, confidence scores
(see Sec.~\ref{subsec:triples}) are used as weights.
Having different confidence scores for S-P and P-O
fits in well in this model as weights for the corresponding edges.
For \textit{alignment edges}, the weight is the similarity between the
respective pair
of entity nodes or 
relation nodes. 
See Sec.~\ref{subsec:corpora} for specifics of these computations.
For \textit{type edges} we set edge weights to
$1.0$,
as these are the most reliable
(relative to the noisier categories).

\section{Graph Algorithm}
\label{sec:graphalgorithm}

\subsection{Computing Group Steiner Trees}
\label{subsec:gst}

\myparagraph{Cornerstones}
To find answers in the quasi KG, we 
first identify pivotal nodes that we call
{\em cornerstones}:
every node that matches a word or phrase in the question,
with similarity
above a threshold,
becomes a cornerstone.
%
For example,
\phrase{FIFA 2018 final} from the example question
is matched by \phrase{2018 FIFA WC Final}
and \phrase{Russia 2018 Final}
(high similarity via lexicons)
in the graph.
Also, relation nodes (e.g.,
\phrase{born} and \phrase{Angolan descent}, with high
word2vec embedding
similarity to \phrase{descent})
and type nodes (e.g., \phrase{French professional footballer} and
\phrase{31-year-old footballer} matching
question term \phrase{footballers})
become cornerstones. All cornerstone nodes for our running example
question
have thick borders
in Fig.~\ref{fig:quasiKG-example-new}.
%
These nodes are weighted based on the matching or
similarity scores.

\myparagraph{Group Steiner Trees} 
The key idea for identifying answer candidates is that these nodes should
be tightly connected to many cornerstones.
To formalize this intuition, we  consider three factors:
(i) answers lie on paths connecting cornerstones,
(ii) short paths are preferred, and
(iii) paths with higher weights are better.
These criteria are 
captured by the notion of a {\bf Steiner Tree}:
%
\squishlist
\item Given an undirected and weighted graph $(V,E)$ with nodes $V$,
edges $E$, and weights $w_{ij} \geq 0$ (for the edge between
nodes $i$ and $j$), and given a subset $T \subseteq V$ of nodes called
{\em terminals},
compute a tree $(V^*,E^*)$
where $V^* \subseteq V, E^* \subseteq E$ that
connects all terminals $T$ and
has minimum cost in terms of total edge weights:
$\min \sum_{ij\in E^*} w_{ij}$ with $T \subseteq V^*$.
\squishend
%
For two terminals, the solution is the shortest path,
but our application comes with many
\textit{terminals, namely the cornerstones}.
Moreover, our terminals are grouped into sets
(the cornerstones
per token of the question),
and it suffices to 
include at least one terminal from each set in the Steiner Tree.
This generalized problem is known as computing a
{\bf Group Steiner Tree (GST)}~\cite{DBLP:journals/jal/GargKR00,
li2016efficient,
ding2007finding}:
\squishlist
\item Given an undirected and weighted graph $(V,E)$ and given groups of
terminal nodes
$\{T_1, \dots, T_l\}$
with each $T_\nu \subseteq V$, compute
the minimum-cost tree $(V^*,E^*)$ that connects 
at least one node from each of
$\{T_1, \dots, T_l\}$:
$min \sum_{ij\in E^*} w_{ij}$ s.t. $T_\nu \cap V^* \neq \phi$,
$\forall T_\nu$.
\squishend
Answer candidates for a question are
\textit{inner nodes} of a GST
(non-terminals).
For example, the quasi KG of Fig.~\ref{fig:quasiKG-example-new} shows
two GSTs, with
nodes in red and blue, respectively, that contain
correct answers \phrase{Samuel Umtiti} and
\phrase{Blaise Matuidi} (underlined).
Algorithms for computing GSTs typically operate on undirected graphs with
non-negative edge weights
reflecting costs. Hence, we cast the quasi KG into an undirected graph by
ignoring the orientation of edges, and we convert the 
$[0,1]$-normalized similarity-score weights
into cost weights by setting $cost = 1 - score$.
Node weights were used for cornerstone selection and are disregarded for
the GST computation.

\myparagraph{Algorithm} 
Steiner trees are among the classical NP-complete problems, 
and this holds for the GST problem too. 
However, the problem has tractable fixed-parameter complexity when the
number of terminals is treated as
a constant~\cite{DBLP:series/txcs/DowneyF13},
and there are also good polynomial-time
approximation algorithms
extensively applied in the area of keyword search over databases
~\cite{li2016efficient,ding2007finding,kacholia2005bidirectional}.
In {\quest}, we build on the exact-solution method
by~\cite{ding2007finding},
which uses
dynamic programming
and has exponential run-time in the 
length of the question
but has $O(n \log n)$ complexity in the graph size.
%
The algorithm works as follows.
Starting from each terminal, trees are iteratively \textit{grown} 
by adding least-cost edges
from their neighborhoods. Trees are periodically
\textit{merged} 
when common vertices are encountered. A priority queue 
(implemented by a Fibonacci heap) holds
all trees in increasing order of
tree costs, and maintains the set of
terminals a specific tree \textit{covers}. 
The process stops when a tree is found that \textit{covers}
all terminals (contains at least one terminal per group) for
the GST problem. 
This bottom-up dynamic programming approach using a priority queue
ensures optimality of the result.

\myparagraph{Relaxation to GST-\textit{k}} 
In our setting, we are actually interested in the top-$k$ trees
in ascending order of
cost~\cite{li2016efficient,ding2007finding}. 
This is for robustness, as the best tree may be a
graph with
cornerstones only;
then we cannot read off any answers from the GST.
Moreover, 
computing several trees provides us with a good way of
ranking multiple candidate answers (see `Answer scoring' below).
Therefore, we use the extended GST-$k$ algorithm
of~\cite{ding2007finding}, 
where $k$ is typically small ($\leq 50$).
The priority-queue-based algorithm 
naturally supports this
top-$k$ computation (fetch top-$k$ trees instead of the top-$1$ only). 



\subsection{Filtering and Ranking Answers}
\label{subsec:answers}

The GSTs may contain several candidate answers, and
so it is crucial to filter and rank the candidates.

\myparagraph{Answer filtering} 
We first remove all candidates that are
not entities (i.e., relation and type nodes are not considered). 
The main pruning is based on lexical {\em type checking}.
For a given question, we infer its expected answer type 
using lexico-syntactic patterns from~\cite{ziegler:17}.
This expected answer type is then compared to the types of
a candidate answer
(type node labels attached to entity candidate),
using 
cosine similarity between 
word2vec
embeddings~\cite{mikolov13distributed}.
Similarity between multi-word phrases is
performed by first averaging individual word vectors
in the phrase~\cite{wieting2016towards},
followed by computing the cosine similarity between
the phrase embeddings.
%
Candidates that do not have types with similarity above
a threshold are dropped. 
%
%

\myparagraph{Answer aggregation} Since answers are surface forms
extracted
from text, we need to reduce redundancy (e.g. to avoid
returning both \phrase{Umtiti} and \phrase{Samuel Umtiti} to the user).
{\quest} aggregates answers based on (i) token sequences, and
(ii) alignments. 
For (i), two answers are merged if one is 
a \textit{subsequence} (not necessarily substring) of another
(e.g., \phrase{Paul Labile Pogba} and \phrase{Paul Pogba}).
For (ii), two answers are merged if
there is an alignment edge between them.
This indicates that they are possibly aliases of the same
entity (e.g., \phrase{Christiano Ronaldo} and \phrase{CR7}).

\myparagraph{Answer scoring}
After aggregation, answers are scored and ranked by exploiting
their presence in
multiple GSTs. 
However, instead of simple counts, we
consider a weighted sum by considering the inverses of the tree cost
as the weight for a GST.
We examine effects of alternatives like 
total node weights
in these GSTs,
and distances of answers to cornerstones,
in our empirical analysis later (Sec.~\ref{sec:detailed}).
%

\section{Evaluation Setup}
\label{sec:setup}

\subsection{Rationale for Experiments}
\label{subsec:rationale}

The key hypotheses that we test in experiments
is that {\quest} can handle {\em complex questions}
that involve multiple entities and relations, and can cope
with the {\em noise in the quasi KG}.
Popular QA benchmarks like
WebQuestions~\cite{DBLP:conf/emnlp/BerantCFL13},
SimpleQuestions~\cite{bordes2015large},
TREC~\cite{agichtein2015overview,dietz2017trec},
QALD~\cite{DBLP:conf/esws/UsbeckNHKRN17},
or SQuAD~\cite{rajpurkar2016squad},
are not suitable, as they mostly focus on answering simple
questions or understanding natural language
passages. 
In contrast, we are interested in computing
direct answers for ad hoc information needs by
advanced users, tapping into all kinds of contents
including informal text and Web tables.
Therefore, we adopted the benchmark
from~\cite{abujabal2017automated}
for structurally complex questions,
and we compiled a new benchmark of complex questions from trending topics
with questions that stress the dynamic and ad hoc nature
of evolving user interests. 
Further, questions in both of these benchmarks
indeed require stitching information across \textit{multiple documents}
for faithful answering, another desideratum for evaluating
the capability of {\quest}.

As for baselines against which we compare {\quest},
we focus on unsupervised and distantly supervised methods.
The latter include neural QA models which are pre-trained
on large
question-answer collections, with additional
input from word embeddings. 
These methods are well-trained for QA in general, but not
biased towards specific benchmark collections.
We are interested in robust behavior for ad hoc questions,
to reflect the rapid evolution and unpredictability
of topics
in questions on the open Web.
Hence this focus on unsupervised and distantly supervised methods.


\vspace*{-0.3cm}
\subsection{Question Benchmarks}
\label{subsec:benchmarks}


\myparagraph{Complex questions from WikiAnswers (CQ-W)} 
This is
a set of $150$ complex fact-centric
questions~\cite{abujabal2017automated}
paired with answers that are 
extracted from a curated WikiAnswers corpus~\cite{fader2013paraphrase}.
Questions in this dataset were specifically
sampled to have multiple
entities and/or relations. 
Unfortunately, baselines in~\cite{abujabal2017automated}
cannot be adopted here,
as they only run over KGs and cannot operate over text. 
%

\myparagraph{Complex questions from Trends (CQ-T)} 
To study situations where KG incompleteness is a major concern,
we created a new benchmark
of $150$ \textit{complex} questions
using {\em emerging entities} from Google Trends,
including entities not having Wikipedia pages at all.
Five students visited \url{https://trends.google.com/trends/}, where
\phrase{trending searches} lists topics of current interest (with USA as
location). 
For every trending topic, the students
looked at \phrase{related queries} 
by Google users on this topic.
Wherever possible, the students then selected
a fact-centric
information need from these queries 
(e.g., \utterance{caitlin mchugh engaged})
and augmented it into a more complex question
(e.g., Q: \utterance{Which TV actress was engaged to John Stamos
and had earlier played in the Vampire Diaries?}; A:
\phrase{Caitlin McHugh}).
The added
nugget of complexity is a mix of \textit{conjunctive, compositional,
	temporal},
and other \textit{constraints}. 
Finally, the students provided answers
by searching the Web.
Each student contributed $30$ (question, answer) pairs.

Wikidata is one of the
most popular KGs today: we found that $29\%$ of questions in CQ-T
\textit{do not have their answer entities in Wikidata}
and $50\%$ \textit{do not have Wikidata facts} connecting question and
answer entities (as of January 2019).
This is often due to relations like (\textit{interviewed, co-appeared
	in event,
married at venue}, etc.) which are of popular interest yet
beyond most KGs.
This illustrates \textbf{KG-incompleteness} and motivates our focus on
answering from dynamically retrieved Web text.

Answers to all $300$ questions 
are manually augmented with \textbf{aliases} ($2.5$ aliases per original
gold answer
on average) 
to be \textit{fair to
competing systems} for extracting correct alternative surface forms
(e.g., \phrase{Cristiano Ronaldo dos Santos Aveiro} is expanded with 
\textit{subsequences} \phrase{Cristiano Ronaldo},
\phrase{Ronaldo}, etc.).
The mean number of unique gold answers 
is about $1.42$
($243/300$ questions have exactly one correct answer).
We verified that
almost all questions
($93\%$ in CQ-W
and
$98\%$ in CQ-T)
require aggregating
\textbf{multi-document evidence}:
there are \textit{no single pages}
in our corpora containing \textit{all} the
information needed to accurately answer them.

\vspace*{-0.1cm}
\subsection{Text Corpora and Quasi KGs}
\label{subsec:corpora}

\begin{table} [t]
	\small
	\setlength{\tabcolsep}{1ex}
\begin{tabular}{l ccc c ccc}
	\toprule
	\multirow{2}{*}{\textbf{Dataset}}
	& \multicolumn{3}{c}{\textbf{\#Nodes}} 					
    &&  \multicolumn{3}{c}{\textbf{\#Edges}}				\\
    \cmidrule{2-4}\cmidrule{6-8}
    & \textbf{Entity} 	& \textbf{Relation} 	& \textbf{Type}     && \textbf{Triple} 	& \textbf{Alignment} 	& \textbf{Type} \\ \midrule
    \textbf{CQ-W} 	
    & $501$ 			& $466$ 				& $28$ 		   	 	&& $1.2$k 			& $5.2$k 				& $434$ 		\\ 
    \textbf{CQ-T} 
	& $472$ 			& $375$ 				& $23$ 		    	&& $1$k 			& $13.2$k 		    	& $436$ 		\\ \bottomrule
\end{tabular}

 \mycaption{Basic properties of quasi KGs, averaged over all questions.}
   \vspace*{-1cm}
  \label{tab:graph-size}  
\end{table}

To decouple system performance from the choice of text corpora,
we experimented
with a number of scenarios for sampling pseudo-relevant documents using
Web search results~\cite{sun2015open}:
\squishlist
\item Top-$10$ documents from Google Web search, where
the whole question
was issued as a keyword query;
\item To weaken the effect of Google's (usually high-quality)
ranking from
the QA performance, 
we constructed different settings
for \textit{stratified sampling}~\cite{voorhees2014effect} of
$10$ documents:
we take the top-$x_1\%$ of documents from the original top-$10$ ranking, 
then sample another $x_2\%$ of our pool randomly from ranks
$(0.1 * x_1 + 1$)
to $25$,
then take the remaining $x_3\%$ from ranks $26$ to $50$
(avoiding duplicates wherever applicable), such that
$x_1 \geq x_2 \geq x_3$.
We use the following five configurations of $x1-x2-x3$, with
gradually ``degrading'' ranking quality:
$60-30-10$ ({\em Strata 1}), $50-40-10$ ({\em Strata 2}), $50-30-20$
({\em Strata 3}),
$40-40-20$ ({\em Strata 4}), and $40-30-30$ ({\em Strata 5}).
\squishend
These varied choices of strata reduce influence of
the underlying search engine.
Stanford CoreNLP~\cite{manning2014stanford} was used for POS tagging
and NER on all 
documents. 

\myparagraph{Similarities and thresholds} Entity similarity scores
for alignment edges
were computed using 
the AIDA dictionary~\cite{hoffart2011robust}.
It consists of a large lexicon
(an updated version was obtained from the authors
of~\cite{hoffart2011robust})
of (entity, mention) pairs, where
a \textit{mention} refers to the surface form in the text (like our
\textit{node labels}), and a canonicalized entity is specified by
its Wikipedia identifier. The similarity between two mentions
is computed
as the Jaccard index of the sets of entities they
refer to. 
%
All other similarities, for relations and types, 
require \textit{soft matching} and are computed using 
cosine similarities between $300$-dimensional word/phrase embeddings
based on  
word2vec~\cite{mikolov13distributed}.
All three thresholds are set to $0.5$:
(i) cornerstone selection, (ii) alignment edge insertion, and 
(iii) answer merging; no tuning is involved.
The number of top-$k$ GSTs to use was 
set to $50$ (effect of this choice is examined later
in Sec.~\ref{sec:detailed}).
Note that $50$ is a very small fraction of
all possible trees in the graph containing the cornerstones
(mean $= 983$, max $\simeq 13$k).
Summary statistics
of our noisy quasi KGs are in Table~\ref{tab:graph-size}. 

\begin{table*} [t]
	\resizebox{\textwidth}{!}{
	\setlength{\tabcolsep}{0.8ex}
\begin{tabular}{l c ccccccc ccccccc}			\toprule
		\multirow{2}{*}{\textbf{Method}}
	&	\multirow{2}{*}{\textbf{Metric}}
	&& 	\multicolumn{6}{c}{\textbf{ComplexQuestions from WikiAnswers (CQ-W)}}			
	&&	\multicolumn{6}{c}{\textbf{ComplexQuestions from Trends (CQ-T)}} \\
	\cmidrule{4-9} \cmidrule{11 - 16}
	& 	&& 	\textbf{Top}	& \textbf{Strata-1} &	\textbf{Strata-2}	& \textbf{Strata-3} & \textbf{Strata-4}	& \textbf{Strata-5} 
	&& 	\textbf{Top}	& \textbf{Strata-1} &	\textbf{Strata-2}	& \textbf{Strata-3} & \textbf{Strata-4}	& \textbf{Strata-5} \\ \midrule
	\quest					
	& 	\multirow{4}{*}{MRR}
	&& $\boldsymbol{0.355}$*	& $\boldsymbol{0.380}$*	& $\boldsymbol{0.340}$*	& $\boldsymbol{0.302}$*	& $\boldsymbol{0.356}$*	& $\boldsymbol{0.318}$*	
	&& $\boldsymbol{0.467}$*	& $\boldsymbol{0.436}$*	& $\boldsymbol{0.426}$*	& $\boldsymbol{0.460}$*	& $\boldsymbol{0.409}$*	& $\boldsymbol{0.384}$* \\
	
	\method{DrQA}~\cite{chen2017reading}
	& 		
	&&	$0.226$					& $0.237$				& $0.257$				& $0.256$				& $0.215$				& $0.248$	
	&& 	$0.355$					& $0.330$				& $0.356$				& $0.369$				& $0.365$				& $0.380$ 				\\
		\bfs~\cite{kasneci2009star}
	& 		
	&& 	$0.249$					& $0.256$				& $0.266$				& $0.212$				& $0.219$ 				& $0.254$	
	&& 	$0.287$					& $0.256$				& $0.265$				& $0.259$				& $0.219$				& $0.201$				\\
	\method{ShortestPaths}					
	& 		
	&& 	$0.240$ 				& $0.261$				& $0.249$				& $0.237$				& $0.259$				& $0.270$
	&& 	$0.266$ 				& $0.224$				& $0.248$				& $0.219$ 				& $0.232$				& $0.222$				\\
	\midrule
	\quest					
	&  \multirow{4}{*}{P@1}	
	&& $\boldsymbol{0.268}$*	& $\boldsymbol{0.315}$	& $\boldsymbol{0.262}$	& $\boldsymbol{0.216}$	& $\boldsymbol{0.258}$*	& $\boldsymbol{0.216}$
	&& $\boldsymbol{0.394}$*	& $\boldsymbol{0.360}$*	& $\boldsymbol{0.347}$*	& $\boldsymbol{0.377}$*	& $\boldsymbol{0.333}$*	& $\boldsymbol{0.288}$ 	\\
	\method{DrQA}~\cite{chen2017reading}
	&  	
	&&	$0.184$					& $0.199$				& $0.221$				& $0.215$				& $0.172$				& $0.200$	
	&& 	$0.286$					& $0.267$				& $0.287$				& $0.300$				& $0.287$				& $0.320$				\\
	\bfs~\cite{kasneci2009star}
	&  
	&& $0.160$ 					& $0.167$				& $0.193$				& $0.113$				& $0.100$				& $0.147$
	&& $0.210$ 					& $0.170$				& $0.180$				& $0.180$				& $0.140$				& $0.130$ 				\\
	\method{ShortestPaths}					
	& 		
	&& 	$0.147$					& $0.173$				& $0.193$				& $0.140$				& $0.147$				& $0.187$
	&& 	$0.190$					& $0.140$				& $0.160$				& $0.160$				& $0.150$				& $0.130$				\\	
	\midrule
	\quest					
	& 	\multirow{4}{*}{Hit@5}
	&& $\boldsymbol{0.376}$		& $\boldsymbol{0.396}$	& $\boldsymbol{0.356}$	& $\boldsymbol{0.344}$	& $\boldsymbol{0.401}$	& $\boldsymbol{0.358}$ 
	&& $\boldsymbol{0.531}$		& $\boldsymbol{0.496}$*	& $\boldsymbol{0.510}$	& $\boldsymbol{0.500}$	& $\boldsymbol{0.503}$	& $\boldsymbol{0.459}$ \\		
	\method{DrQA}~\cite{chen2017reading}
	& 	
	&&	$0.313$					& $0.315$				& $0.322$				& $0.322$				& $0.303$				& $0.340$	
	&&	$0.453$					& $0.440$				& $0.473$				& $0.487$				& $0.480$				& $0.480$				\\
	\bfs~\cite{kasneci2009star}
	& 	
	&& 	$0.360$					& $0.353$				& $0.347$				& $0.327$				& $0.327$				& $0.360$				
	&&  $0.380$					& $0.360$				& $0.370$				& $0.360$				& $0.310$				& $0.320$				\\
	\method{ShortestPaths}					
	& 	
	&& 	$0.347$					& $0.367$				& $0.387$				& $0.327$				& $0.393$				& $0.340$	
	&&	$0.350$					& $0.320$				& $0.340$				& $0.310$				& $0.330$				& $0.290$				\\	
	\bottomrule	
\end{tabular}

}
	\mycaption{Performance comparison 
of methods
on top-10 and stratified
	search results from the Web. For every
	metric, the best value per column is in \textbf{bold}.
	``$*$'' denotes statistical significance of {\quest} over
	\method{DrQA},
with $p$-value $\le 0.05$ for a one-tailed paired $t$-test.}
	\label{tab:graph-qa}
	\vspace*{-0.7cm}
\end{table*}

\subsection{Baselines and Metrics}
\label{subsec:baselines}


\myparagraph{Neural QA}
As a strong neural baseline, we select 
\method{DrQA}~\cite{chen2017reading},
a very recent open-source QA system. 
\method{DrQA} has large-scale training on
SQuAD~\cite{rajpurkar2016squad},
and is based on
recurrent neural networks and multitask learning.
It was designed for reading comprehension, but it can select 
relevant documents from a corpus and extract
the best answer span from these documents.
\method{DrQA} and other baselines run on the same set of input documents 
that {\quest}
is exposed to.
Specifically, \method{DrQA} has two components DocumentRetriever and 
DocumentReader,
which are both run on the top-$10$ and stratified corpora from
Google Web search.



\myparagraph{Graph-based algorithms} 
As a competitor to the GST algorithm of {\quest},
we adapted the \textit{breadth-first search} ({\bfs}) phase
of the STAR algorithm~\cite{kasneci2009star}
for entity relatedness in curated KGs.
The full STAR method
can only work in the presence of a taxonomic backbone in the KG,
which is inapplicable in our case. 
The {\bfs} baseline
runs graph-traversal
iterators from each terminal node, invoked 
in a
round-robin manner. As soon as the iterators meet, a result
is constructed.
To respect sets of cornerstones, we require only one iterator
from each
group of terminals to meet for a candidate answer. Results are
ranked in 
descending order of their weighted distances from the cornerstones
in the {\bfs} tree.

To examine the importance of the optimal 
subgraph identified by the GST,
we also compare results using \textit{shortest paths} as follows.
We compute
shortest paths in the graph between every pair of terminals,
where each node
in a pair is from a different cornerstone group. Every non-terminal
that lies on any shortest path is a candidate answer. 
An answer is scored by the numbers of different shortest paths
that it lies on, and ranked in descending order of these scores. 
For both {\bfs} and {\shortestPath}, \textit{for fairness, answers are
post-processed by the same type-based filtering and aggregation}
as in {\quest},
before applying respective answer ranking strategies.

\myparagraph{Metrics} We use the Mean Reciprocal Rank (MRR) as the main 
metric. 
We also report other key metrics for QA: Precision@1 (P@1),
which measures
the fraction of times
a correct answer was obtained at rank $1$, and Hit@5, which is $1$ when
one of the top-$5$ results is a gold answer and $0$ otherwise.

\section{Main Results and Insights}
\label{sec:results}

We present our main results in Table~\ref{tab:graph-qa},
and discuss key insights from these comparisons.
Our main experiments test the
postulates:
\squishlist
\item {\quest} outperforms its neural and graph baselines;
\item Performance of {\quest} is robust to corpus perturbations;
\item Multi-document evidence is vital for retrieving correct answers;
\item Group Steiner Trees are key to locating answers in quasi KGs.
\squishend

\myparagraph{Systematic improvement over state-of-the-art} Looking at the
``Top'' and ``Strata'' columns (Sec.~\ref{subsec:corpora}) for both
benchmarks,
we find that {\quest} significantly
and consistently outperforms the neural baseline \method{DrQA}, and other 
graph-based methods, at almost all settings. This performance of {\quest}
is clearly robust to variations in the underlying corpus. We attribute the 
success
of the proposed method to its unique ability to stitch facts from more
than one source, and the powerful GST algorithm that discovers answers in
the large and very noisy quasi KGs.
The task of fetching crisp text answers to complex questions directly
over Web corpora is generally a very difficult
one; this is reflected by relatively low values of MRR (best numbers of
$0.355$ for CQ-W
and $0.467$ for CQ-T
in top-$10$ corpora).

\myparagraph{Effect of corpus variations} Exact reliance on 
Google's top-$10$ Web search results is not
a prerequisite: we show this by weakening the 
search ranking with stratified sampling (as discussed in
Sec.~\ref{subsec:corpora}) by intentionally introducing controlled amounts
of noisy pages in the pseudorelevant corpora. Results are 
in 
columns
labeled ``Strata-1'' through ``Strata-5'' 
in 
Table~\ref{tab:graph-qa}. The key observation is the across-the-board
superiority of {\quest},
and that it was able to cope well with this
injected noise. Note that Google's ranking may not always be perfect, as
the stratified configuration $60-30-10$ (Strata 1) resulted in slightly
better performance than the top-$10$ (e. g. MRR of $0.380$ vs. $0.355$
on CQ-W for {\quest}). We also experimented with the setting
where search was restricted to Google News,
and observed similar trends ($0.261$ MRR for {\quest} vs. $0.227$
for \method{DrQA} on top-$10$, aggregated over CQ-W and CQ-T).
Google search over Wikipedia only
turned out in favor of DrQA ($0.244$ MRR vs. $0.189$ for {\quest},
top-$10$). This is due to the low redundancy of facts
in Wikipedia,
that hurts {\quest} (explained shortly), and the Wikipedia-specific
training of \method{DrQA}.

\begin{table} [t]
	\small
	\setlength{\tabcolsep}{0.5ex}
\begin{tabular}{cc ccc ccc} \toprule
	\textbf{Benchmark} &&& \multicolumn{2}{c}{\textbf{CQ-W}}			&& \multicolumn{2}{c}{\textbf{CQ-T}}	 			\\
	\cmidrule{1-2} \cmidrule{4-6} \cmidrule{7-8}
	\multirow{2}{*}{\textbf{GST Ranks}}		&&& \textbf{Avg. \#Docs}	& \textbf{\#Q's with} 		&& \textbf{Avg. \#Docs}		&\textbf{\#Q's with} \\
											&&& \textbf{in GST}			& \textbf{A in GST} 		&& \textbf{in GST}			& \textbf{A in GST} \\
	\toprule
	$01 - 10$								&&& $2.637$ 				& $48$						&& $3.139$					& $52$ 					\\
	$11 - 20$								&&& $2.789$ 				& $56$						&& $3.156$					& $53$ 					\\
	$21 - 30$								&&& $2.778$ 				& $54$						&& $3.245$					& $55$ 					\\
	$31 - 40$								&&& $2.833$					& $54$						&& $3.267$					& $54$ 					\\
	$41 - 50$								&&& $2.882$					& $57$						&& $3.278$					& $51$ 					\\ \midrule
	\multirow{2}{*}{\textbf{\#Docs in GST}}	&&& \textbf{Avg. Rank}		& \textbf{\#Q's with} 		&& \textbf{Avg. Rank}		& \textbf{\#Q's with} \\
											&&& \textbf{of GST}			& \textbf{A in GST} 		&& \textbf{of GST}			& \textbf{A in GST} \\
	\toprule
	$1$										&&& $24.525$ 				& $12$						&& $22.124$					& $7$ 					\\
	$2$										&&& $27.003$ 				& $37$						&& $24.216$					& $23$ 					\\
	$3$										&&& $25.777$ 				& $53$						&& $27.165$					& $50$ 					\\
	$4$										&&& $27.064$				& $36$						&& $27.069$					& $49$ 					\\
	$5$										&&& $29.291$				& $25$						&& $26.554$					& $29$ 					\\ \bottomrule
\end{tabular}
	\mycaption{Effect of multi-document evidence
		shown via edge contributions by distinct documents to GSTs (on
		top-$10$ corpora). }
	\label{tab:mdev}
	\vspace*{-1cm}
\end{table}

\myparagraph{Usage of multi-document evidence} {\quest} improved over
\method{DrQA} on both benchmarks (MRR of $0.355$ vs. $0.226$ on CQ-W;
$0.467$ vs. $0.355$ on CQ-T, top-$10$), even
though the latter is a supervised deep learning method
trained on the large SQuAD dataset.
This is because
reading comprehension (RC) QA systems
search for the best answer span within a passage, and will not
work well unless the passage
matches the question tokens and contains the answer.
While \method{DrQA}
can additionally select a good set of documents from a collection,
it still relies
on the best document to extract the answer from. {\quest},
by joining fragments of evidence across documents via GSTs, thus improves
over
\method{DrQA} without any training or 
tuning.
{\quest} benefits from multi-document evidence in two ways:
\squishlist
\item Confidence in an answer increases when all conditions for correctness
are indeed satisfiable (and found) only when looking at multiple documents.
This increases
the answer's likelihood of appearing in \textit{some GST}.
\item Confidence in an answer increases when it is spotted
in multiple documents. This increases its likelihood of appearing in
the \textit{top-k} GSTs, as presence in multiple documents increases
weights
and lowers costs of the corresponding edges.
\squishend
A detailed investigation of the use of multi-document information is
presented in Table~\ref{tab:mdev}. We make the following observations:
(i) Looking at the ``Avg. \#Docs in GST'' columns in the upper half,
we see that 
considering the top-$50$ GSTs is worthwhile as all the bins combine
evidence from multiple ($2+$ on average) documents.
This is measured by labeling edges in GSTs with
documents (identifiers) that contribute the corresponding edges.
(ii) Moreover, they also contain
the correct answer uniformly often
(corresponding
``\#Q's with A in GST'' columns; $48-57$ for CQ-W, $51-55$ for CQ-T).
(iii) The bottom half of the table inspects
the inverse phenomenon, and finds that
considering only the top few GSTs is not sufficient
for aggregating multi-document evidence.
(iv) Finally,
there is a sweet spot for GSTs aggregating nuggets from multiple documents
to contain correct answers, and this turns out to be around three
documents (see corresponding ``\#Q's with A in GST'' columns).
This, however, is an effect of our questions
in our benchmarks,
that are not complex enough to require stitching evidence
across more than three documents.
%

Deep-learning-based RC methods over text can handle syntactic complexity
very well, but are typically restricted to identifying answer spans from
a single
text passage. \method{DrQA} could not properly tap the answer evidence that
comes from combining cues spread across multiple documents.


\myparagraph{Impact of Steiner Trees}
Observing the graph-based methods,
{\quest} is clearly better than both
{\bfs} and {\shortestPath}, obtaining correct answers at better ranks. 
This demonstrates that computing cost-optimal GSTs
is indeed crucial, and cannot be easily approximated by simpler methods.
It is not simply the connectivity alone that qualifies a node as a good answer,
which is why the simpler {\shortestPath} method substantially loses against
{\quest}.
Computing the GST can be viewed as a \textit{joint disambiguation}
for all
the semantic items in the question, like entities and predicates in a KG.

\begin{table} [t]
	\small
	\newcommand{\tabent}[1]{\begin{minipage}[c]{8.3cm}%
		\raggedright#1\\[-2ex]~\end{minipage}}
\setlength{\tabcolsep}{0.5ex}
\begin{tabular}{l}
	\toprule
\multicolumn{1}{l}{\textbf{Questions from CQ-W}  ($P@1 = 1$)} \\ \midrule
\tabent{\textbf{Q}:\utterance{which actor is married to kaaren verne and played in casablanca?}\\
\textbf{A}: \phrase{Peter Lorre}}
\\[-1ex]
\tabent{\textbf{Q:}\utterance{what river flows through washington and oregon?} \\
\textbf{A:} \phrase{Columbia River}} 
\\[-1ex]
\tabent{\textbf{Q}: \utterance{what movie did russell crowe and denzel washington 
work on together?} 
\\
\textbf{A}: \phrase{American Gangster}} 
\\[-2ex]
\midrule
\multicolumn{1}{l}{\textbf{Questions from CQ-T}  ($P@1 = 1$)} \\ \midrule
\tabent{\textbf{Q}: \utterance{where did sylvie vartan meet her future husband 
johnny hallyday?}\\
\textbf{A}: \phrase{Paris Olympia Hall}}
\\[-1ex]
\tabent{\textbf{Q:} \utterance{which aspiring model split with chloe moretz and is 
dating lexi wood?} \\
\textbf{A:} \phrase{Brooklyn Beckham}}
\\[-1ex]
 \tabent{
 	\textbf{Q:}\utterance{which japanese baseball player was contracted for los angeles 
 		angels who also played for  hokkaido 	nippon-ham fighters?}\\
 	\textbf{A:} \phrase{Shohei Ohtani}
 }
 \\[-2ex]
 \bottomrule
\end{tabular}

%
	\mycaption{Examples of correctly answered questions by {\quest}
		but not by any of the baselines (on top-10 corpora).} 
	\label{tab:anecdotes}
	\vspace*{-0.9cm}
\end{table}

\myparagraph{Anecdotal results} 
To highlight the complexity of questions that are within
our reach, 
Table~\ref{tab:anecdotes} shows representatives 
where {\quest}
had the correct answer at the very first rank, but all
the baselines failed. 
Note that some of the questions refer to long-tail entities not covered yet
by KGs like
Wikidata, and also have predicates like \textit{met,
split}, and \textit{dated}, which are beyond the scope of KGs.

\section{Analysis and Discussion}
\label{sec:detailed}



\begin{table*} [t]	
	\resizebox*{\textwidth}{!}{
	\begin{tabular}{l c c | l c c | l c c}
		\toprule
		\textbf{Graph configuration}	&	\textbf{CQ-W}			&	\textbf{CQ-T}			& \textbf{Answer ranking criterion}			&	\textbf{CQ-W}			&	\textbf{CQ-T}			& \textbf{Error scenario}						& \textbf{CQ-W}	&	\textbf{CQ-T}	\\ \toprule 
		Full configuration				&	$\boldsymbol{0.355}$	&	$\boldsymbol{0.467}$	& Wted. sum of GSTs (inv. tree cost sum)	&	$\boldsymbol{0.355}$	&	$\boldsymbol{0.467}$	& Ans. not in corpus							& $1\%$			&	$7\%$			\\ 
		No types						&	$0.321$					&	$0.384$*				& Wted. sum of GSTs (node wt. sum)			&	$0.318$*				&	$0.426$*				& Ans. in corpus but not in quasi KG			& $23\%$		&	$30\%$			\\
		Degenerate edge weights			&	$0.282$*				&	$0.377$*				& Count of GSTs								&	$0.334$					&	$0.432$*				& Ans. in quasi KG but not in top-$50$ GSTs		& $10\%$		&	$6\%$			\\ 
		No entity alignment				&	$0.329$*				&	$0.413$*				& Wted. dist. to cornerstones				&	$0.321$*				&	$0.417$*				& Ans. in top-$50$ GSTs but not in candidates	& $1\%$			&	$7\%$			\\
		No predicate alignment			&	$0.337$					&	$0.403$*				& Unwted. dist. to cornerstones				&	$0.257$*				&	$0.372$*				& Ans. in candidates but not in top-$5$			& $66\%$		&	$49\%$			\\ \bottomrule
	\end{tabular}}
	\mycaption{Understanding {\quest}'s mechanism with top-$10$ corpora.
	\textbf{Left:} Graph ablation (MRR);
	\textbf{Middle:} Answer ranking (MRR);
	\textbf{Right:} Error analysis (Hit@5 = 0). Highest column values in the first
	two sections in \textbf{bold}. Significant drops from these
	values are shown with *.}
	\label{tab:analysis}
	\vspace*{-0.5cm}
\end{table*}

\myparagraph{Graph ablation experiments} 
The noisy graph is the backbone of
{\quest} and has several components working in tandem. We systematically
analyzed this interplay of components by deactivating each separately
(Table~\ref{tab:analysis}).
%
The key insight is that type nodes and edges are
essential to the success of {\quest}; removing them results in
degraded performance. 
Next, using informative edge weights
driven by document proximity and alignment levels, are another vital element
(MRR drops with degenerate edge weights).
Removing alignment edges also adversely affects performance.


\myparagraph{Answer ranking variants} 
Our answer ranking strategy
is motivated by
considering a weighted (with reciprocal of tree cost) sum for exploiting
answer presence
in multiple GSTs.
Nevertheless, we explored various
alternatives to this choice, and observed (Table~\ref{tab:analysis}):
(i) just counting GSTs in which an answer is present is not enough;
(ii) reusing node weights 
for scoring trees
with answers
does not really help;
(iii) There is no additional
benefit in zooming in to consider the position of an answer 
\textit{within a GST};
nodes that are the closest to cornerstones are not
necessarily the best answers.
However, differences between the first four choices are not
very high: hence, {\quest} is robust to slight
ranking variants as long as answer evidence across multiple GSTs
is considered.

%

\myparagraph{Error analysis}
Failure cases for {\quest}, and corresponding occurrences
are shown in Table~\ref{tab:analysis}. We
treat a case as an error when {\quest} cannot locate any
correct answer in the top-$5$ (Hit@5$= 0$).
The first case suggests use of a better retrieval model,
considering semantic matches.
The second is a key reason for failure, and demands 
an Open IE extractor with better fact recall.
The third scenario is mostly caused by
matching wrong groups of cornerstones. For example, matching
relation \phrase{born} in question with \phrase{lived}
in the quasi KG; or \phrase{play} (drama) intended as a noun, matching
a relation node \phrase{played} (role).
This can be
improved with better NER, lexicons, and similarity functions.
Case (iv) happens due to pruning by type mismatch.
This calls for more informed prediction and matching of expected answer types,
and improving type extraction from documents.
Situation (v) indicates that improving the ranking function
is the most worthwhile effort for improving performance.

\myparagraph{Effect of Open IE} {\quest}'s noisy triple extractor
results in quasi KGs that contain the correct answer $85.2\%$ of
the times for CQ-W ($82.3\%$ for CQ-T). If we use triples
from Stanford OpenIE~\cite{angeli2015leveraging} to build the
quasi KG instead, the answer is found only $46.7\%$ and $35.3\%$
times for CQ-W and CQ-T, respectively. Thus, in the context
of QA, losing information with precision-oriented, fewer triples,
definitely hurts more than the adding potentially many noisy ones.

\myparagraph{Answering simple questions} On the popular
WebQuestions benchmark~\cite{DBLP:conf/emnlp/BerantCFL13},
{\quest} was able to find a correct answer $54\%$ of the time,
which is respectable in comparison
to the best QA systems specifically trained for this setting
of simple questions. GSTs are augmented with $1$-hop neighbors
of terminals
to handle two-cornerstone (single-entity single-relation) questions.

\begin{figure*} [t]
	\input{fig-threshold.tex}
	        \\[-2ex]
	\mycaption{Robustness of {\quest} to various system configuration
	parameters, on CQ-W with top-$10$ corpora (similar trends on CQ-T).}
	\label{fig:parameters}
	\vspace*{-0.25cm}
\end{figure*}

\myparagraph{Effect of threshold variations}
There are three parameters in 
{\quest}: Number of GSTs $k$, the alignment edge insertion threshold on
node-node similarity, and
cornerstone selection thresholds on node weight.
Variation of
these parameters are shown with MRR at cut-off ranks $r = 1, 3, 5$
in Fig.~\ref{fig:parameters}.
We observe that:
(i) going beyond
the chosen value of $k=50$ 
gives only diminishing returns;
(ii)
Fig.~\ref{fig:align}
shows that
having several alignment edges in the graph
(corresponding to a very low threshold),
actually helps improve
performance though apparently inviting noise;
(iii) {\quest} is not really sensitive to cornerstone
selection thresholds --
the dark zone, indicating good performance, is towards
the interiors of the grid, but broadly spread out:
so most choices of non-extreme
thresholds will work out fine.
The white zone in the top right corner 
corresponds to 
setting both thresholds to very high values
(no cornerstones chosen, resulting in zero performance).


\myparagraph{Run-time} {\quest} computes answers with interactive
response times: 
the median run-time for computing GSTs
was $1.5$
seconds (with a mean of 
about five seconds). 
All experiments were performed with Java 8 on a Linux machine (RHEL-$v6.3$)
using an Intel Xeon E5 CPU
with $256$ GB RAM.


\section{Related Work}
\label{sec:related}
\balance

\myparagraph{QA over text} 
Classical approaches~\cite{ravichandran2002learning}
extracted answers from passages 
that matched most cue words from the
question
followed by statistical scoring.
%
TREC ran a QA benchmarking series from 1999 to 2007,
recently revived it as the LiveQA track. 
%
%
IBM Watson~\cite{watson2012}
extended this
paradigm by combining it with learned models
for special question types.

\myparagraph{QA over KGs}
The advent of large knowledge graphs
like Freebase~\cite{bollacker2008freebase},
YAGO~\cite{suchanek2007yago},
DBpedia~\cite{auer2007dbpedia} and 
Wikidata~\cite{vrandevcic2014wikidata}
has given rise to 
QA over KGs 
(overviews
in~\cite{DBLP:conf/rweb/UngerFC14,DBLP:journals/kais/DiefenbachLSM18}).
The goal is to translate
a natural language question
into a structured query, 
typically in the Semantic Web
language SPARQL, 
that directly operates
on the entities and predicates of the underlying KG
~\cite{DBLP:conf/rweb/UngerFC14,DBLP:journals/kais/DiefenbachLSM18}.
Early work on KG-QA built on paraphrase-based
mappings and query templates that cover simple 
forms of questions that
involve a single entity predicate
~\cite{DBLP:conf/emnlp/BerantCFL13,
DBLP:conf/acl/CaiY13,fader2013paraphrase,
DBLP:conf/www/UngerBLNGC12,DBLP:conf/emnlp/YahyaBERTW12}. 
This line was further advanced 
by~\cite{fader2014open,bast2015more,DBLP:conf/coling/BaoDYZZ16,
abujabal2017automated,DBLP:journals/tkde/Hu0YWZ18},
including the learning of templates from graph patterns in the KG.
However, reliance on templates prevents
such approaches from robustly coping with
arbitrary
syntactic formulations.
This has motivated deep learning
methods with CNNs and LSTMs~\cite{DBLP:conf/emnlp/IyyerBCSD14,
DBLP:conf/acl/YihCHG15,DBLP:conf/acl/DongWZX15,xu2016question,
DBLP:journals/taslp/TanWZYDLZ18}.
%
These have been 
most successful on benchmarks like 
WebQuestions~\cite{DBLP:conf/emnlp/BerantCFL13} and
QALD~\cite{DBLP:conf/esws/UsbeckNHKRN17}.
However, all these methods critically build
on sufficient amounts of training data
in the form of question-answer pairs.
In contrast, {\quest}
is fully unsupervised
and neither needs templates nor training data.


\myparagraph{QA over hybrid sources}
%
Limitations of QA over KGs has 
led to a revival of considering textual sources,
in combination with KGs~\cite{savenkov2016knowledge,xu2016question,sun2018open}.
Some methods like {\paralex}~\cite{fader2013paraphrase}
and \method{OQA}~\cite{fader2014open}
supported noisy KGs
in the form of triple spaces compiled 
via 
Open IE
~\cite{mausam2016open,
DBLP:conf/emnlp/GashteovskiGC17} 
on Wikipedia articles or Web corpora. 
\method{TupleInf}~\cite{khot2017answering}
extended and generalized
{\paralex} 
to complex questions, but is 
limited to \textit{multiple-choice answer options}
and is thus inapplicable for our task.
%
%
\method{TAQA}~\cite{DBLP:conf/cikm/YinDKBZ15}
is another generalization
of 
Open-IE-based QA, by constructing 
a KG of $n$-tuples 
from Wikipedia full-text and 
question-specific search results. 
Unfortunately this method is restricted to questions with
prepositional and adverbial constraints only. 
%
%
%
%
%
%
%
\cite{TalmorBerant:NAACL2018} addressed
complex questions by decomposing them 
into a sequence of simple questions,
but relies on training data obtained via Amazon Mechanical Turk.
%
%
Some methods
start with KGs as a source for candidate
answers and use text corpora 
like 
Wikipedia or ClueWeb 
as additional 
evidence~\cite{xu2016question,DBLP:conf/acl/DasZRM17,sun2018open}, or
start with answer sentences from text
corpora and combine these with KGs
for 
entity answers~\cite{sun2015open,savenkov2016knowledge}. 
Most of these 
are based on
neural networks, and are only designed for
\textit{simple} questions like those in the WebQuestions,
SimpleQuestions, or WikiMovies
benchmarks.
In contrast,
{\quest} can handle
arbitrary kinds of complex questions
and 
can construct \emph{explanatory evidence}
for its answers -- an unsolved concern
for neural methods.

%

%
\myparagraph{Reading comprehension} This is a QA variation
where a question needs to be answered from
a given text
paragraph~\cite{rajpurkar2016squad,DBLP:conf/acl/JoshiCWZ17}.
This is different from the fact-centric answer-finding task considered
here, 
with input from
dynamically retrieved documents. Nevertheless, we compared
with, and outperformed, the state-of-the-art
system \method{DrQA}~\cite{chen2017reading},
which can both select relevant documents and extract answers from them.
Traditional fact-centric QA over text,
and multi-document reading comprehension 
are recently emerging as a joint topic referred to as open-domain question 
answering~\cite{lin2018denoising,dehghani2019learning}.

\section{Conclusion}
\label{sec:conclusion}

%
%
%
%
%
%
%

We presented {\quest}, an unsupervised method for 
QA over dynamically retrieved text corpora
based on Group Steiner Trees.
{\quest} substantially 
outperforms \method{DrQA}, a strong deep learning baseline,
on challenging benchmarks.
As noisy content is unavoidable with Web content
and ad hoc questions for which extensive training
is infeasible, 
{\quest} 
deliberately allows noise
in its 
computational pipeline,
and copes with it using cross-document evidence,
smart answer detection, and
graph-based ranking strategies.
%
%
Adapting QUEST to work
over combinations of text-based quasi KGs and curated KGs
will be the focus of future studies.



\bibliographystyle{ACM-Reference-Format}
\bibliography{strings-shrt,quest}

\end{document}